\documentclass[a4paper,11pt]{article}
\usepackage{graphicx,amssymb,bm,latexsym,color,epsf}
\usepackage{hyperref}
\usepackage{cite}
\makeatletter
\@addtoreset{equation}{section}

\makeatother
\newcommand{\be}{\begin{equation}}
\newcommand{\ee}{\end{equation}}
\newcommand{\bea}{\begin{eqnarray}}
\newcommand{\eea}{\end{eqnarray}}
\newcommand{\vs}[1]{\vspace{#1 mm}}

\newcommand{\cC}{{\cal C}}
\newcommand{\cD}{{\cal D}}
\newcommand{\cH}{{\cal H}}

\newcommand{\cL}{{\cal L}}

\newcommand{\ud}{\mathrm{d}}

\newcommand{\bg}{\bar g}

\newcommand{\Lie}{{\cal L}}
\newcommand{\diff}{\mathit{Diff}}
\newcommand{\sdiff}{\mathit{SDiff}}

\newcommand{\wtdiff}{\mathit{WT\!\diff}}

\begin{document}

\begin{center}
{\Large\bf Gravity with more or less gauging}
\vs{8}

{\large
Steffen Gielen$^{a,}$\footnote{e-mail address: steffen.gielen@nottingham.ac.uk},
Rodrigo de Le\'on Ard\'on$^{b,c,}$\footnote{e-mail address: rdeleon@sissa.it}
and Roberto Percacci$^{b,c,}$\footnote{e-mail address: percacci@sissa.it}
} \\
\vs{8}
$^a${\em School of Mathematical Sciences, University of Nottingham, Nottingham NG7 2RD, UK}

$^b${\em International School for Advanced Studies, via Bonomea 265, 34136 Trieste, Italy}

$^c${\em INFN, Sezione di Trieste, Italy}
\end{center}
\vs{12}

\begin{abstract}
General Relativity is usually formulated as a theory with
gauge invariance under the diffeomorphism group,
but there is a `dilaton' formulation where it is in addition invariant under Weyl transformations, and a `unimodular' formulation
where it is only invariant under the smaller group of 
special diffeomorphisms.
Other formulations with the same number of gauge generators,
but a different gauge algebra, also exist.
These different formulations provide examples of 
what we call `inessential gauge invariance',
`symmetry trading' and `linking theories';
they are locally equivalent,
but may differ when global properties of the solutions
are considered.
We discuss these notions in the Lagrangian
and Hamiltonian formalism.
\end{abstract}

\vs{6}

\section{A circle of theories}
\label{introsec}

Physical systems often admit different descriptions.
We shall be concerned with a special aspect of this,
namely gauge theories which can be described as having
larger or smaller gauge groups.
The question arises both in gravity and in Yang-Mills
theories.
We will only consider one particular example, namely
four different formulations of GR,
where by GR we mean the theory of gravity
based on Einstein's equations.
Let us begin by listing these four formulations,
then we shall discuss reasons to prefer one over another.

\subsection{EG}

In Einstein's standard formulation (which we call Einstein Gravity, 
or EG), gravity is described
by the dynamics of a metric $g_{\mu\nu}$ and the action is
\be
S_{{\rm EG}}(g)=Z_N\int \ud^dx\, \sqrt{|g|}\, R
\qquad\mathrm{where}\qquad Z_N=\frac{1}{16\pi G}\ .
\label{actionE}
\ee
In the four-dimensional case $Z_N$ is one half of the squared reduced Planck mass $M_P^2 = (8\pi G)^{-1}$.

\subsection{DG}

We will call (conformal) Dilaton Gravity (DG) a reformulation of GR
where in addition to the metric there is a scalar field $\phi$
and the action is defined by
$S_{{\rm DG}}(g,\phi)=S_{{\rm EG}}(\bg)$
where
\be
\bg_{\mu\nu}=\left(\frac{\alpha}{Z_N}\right)^\frac{2}{d-2}
\phi^2 g_{\mu\nu}
\label{substitute}
\ee
and $\alpha$ is a pure number.
This action had been considered by Dirac in \cite{dirac}, 
motivated by Weyl's idea of allowing arbitrary units 
of length at each spacetime point\cite{weyl}.
For this reason one could also read DG as `Dirac Gravity'.

Using the Weyl transformation properties of the curvature,
the action becomes
\be
S_{{\rm DG}}(g,\phi)=\alpha\int \ud^dx\, \sqrt{|g|}\, 
\phi^{d-2}\left[R
-2(d-1)\phi^{-1}\nabla^2\phi
-(d-1)(d-4)\phi^{-2}(\nabla\phi)^2\right]\ .
\nonumber
\label{actionD1}
\ee
By construction, this action is automatically invariant under the Weyl transformations
\be
g_{\mu\nu}\to\Omega^2 g_{\mu\nu}\ ,
\qquad
\phi\to\Omega^{-1}\phi\ .
\label{fakeweyl}
\ee
For $d=4$, (\ref{actionD1}) is the usual action for a conformally coupled scalar.
In other dimensions, it is more convenient to define 
$\phi=\psi^\frac{2}{d-2}$,
so that (\ref{actionD1}) can be rewritten
\be
S'_{{\rm DG}}(g,\psi)=\alpha\int \ud^dx\, \sqrt{|g|}\, 
\left[\psi^2R
-4\frac{d-1}{d-2}
\psi\nabla^2\psi
\right]\ .
\label{actionD2}
\ee
In this case Weyl transformations
act by $\psi\to\Omega^{-\frac{d-2}{2}}\psi$.

These actions are properly Weyl-invariant,
not just invariant up to total derivatives.
They have the right sign if we think of them as actions for gravity, 
but the wrong sign if we think of them as actions of a scalar field.
In the latter case it is customary to normalise the kinetic term, leading to the choice
\be
\alpha=\frac{d-2}{8(d-1)}\ .
\label{alpha}
\ee
Also, in this case one usually integrates by parts
the kinetic term.
If we do this and neglect the total derivatives,
then the action is only invariant up to a total derivative.
In the following we will assume that such integrations can always be done, i.e.~we will discard all boundary terms.

\subsection{UG}

Unimodular Gravity (UG) \cite{Anderson:1971pn,vanderBij:1981ym,Unruh:1988in,Buchmuller:1988wx,Ellis:2010uc}
is a theory where the determinant of the
metric is fixed. Usually one just chooses $\sqrt{|g|}=1$.
This condition, however, restricts the allowed coordinate systems
and furthermore is incompatible with the choice of
dimensionless coordinates that we find to be often desirable.
We thus define UG by the condition
\be
\sqrt{|g|}=\omega\ ,
\label{constraint}
\ee
where $\omega$ is a fixed scalar density of weight one.
Locally any scalar density can be transformed into any other
by a diffeomorphism, so (\ref{constraint}) is almost
entirely a gauge statement.
However, the total four-volume (when finite) is $\diff$-invariant
and hence in principle observable,
so the statement 
$$
V_{\sqrt{|g|}}(M)=V_\omega(M)\ ,
$$
which is obtained by integrating (\ref{constraint}) over the spacetime manifold $M$,
is a genuine physical constraint. This global difference between EG and UG leads to a different role of the cosmological constant, which appears in UG as an integration constant for each solution, rather than a fixed parameter in the action.

Since $\omega$ is a fixed non-dynamical background,
it breaks the gauge invariance 
to the group $\sdiff$ of volume-preserving,
or `special' diffeomorphisms.

In the `minimal' formulation, one thinks of UG
as EG where one has partly fixed
the $\diff$ invariance by the condition (\ref{constraint}).
The action of minimal UG is simply
\be
S_{{\rm UG}}(g)=Z_N\int \ud^dx\, \omega\, R\ .
\label{actionUG}
\ee

We note that for the linearised theory,
or in applications of the background field method,
the constraint (\ref{constraint})
is automatically enforced by writing
\be
\label{exppar}
g_{\mu\nu}=\bg_{\mu\rho}(e^X)^\rho{}_\nu\ ,
\ee
where $\sqrt{|\bg|}=\omega$ and demanding that $X^\rho{}_\nu$ 
be traceless \cite{Ardon:2017atk}.

\subsection{UD}

We can also make DG unimodular by imposing (\ref{constraint}) on the metric $g$ in (\ref{actionD2}). 
This amounts to a partial gauge fixing of both diffeomorphisms and Weyl transformations, which both change the determinant of the metric. The symmetry group $\mathit{Diff}\ltimes\mathit{Weyl}$ of DG is broken to its `special' subgroup; 
any further diffeomorphism that would change the metric determinant must be compensated by a Weyl transformation in order to preserve (\ref{constraint}).
We call this Unimodular Dilaton gravity (UD).
It has been discussed in \cite{Katanaev:2005xd}.

The action of UD is
\be
S_{{\rm UD}}(g,\psi)=\alpha\int \ud^dx\, \omega\, 
\left[\psi^2R
-4\frac{d-1}{d-2}
\psi\nabla^2\psi
\right]\,.
\label{actionUD}
\ee
There is no analogue of the constraint on the total spacetime volume in this case: the volume measured with the metric $g$ is gauge-dependent in DG, due to the Weyl symmetry, and the volume measured with the metric $\bar{g}$ is unaffected by the unimodularity constraint.

The unimodular condition has been employed in applications of DG
to cosmology, where one can use it to shift the expansion of the Universe from the metric determinant into the scalar field $\psi$; we will discuss cosmological applications in section \ref{discsec}.

We could write  $\mathit{S(Diff}\ltimes\mathit{Weyl)}$ for the symmetry group of UD, but the group is actually just the diffeomorphism group $\diff$ acting on the fields in a non-standard way. The usual action of $\diff$ is generated by Lie derivatives, 
\be
\delta_\epsilon g_{\mu\nu} = \Lie_\epsilon g_{\mu\nu} = \nabla_\mu\epsilon_{\nu}+\nabla_\nu\epsilon_{\mu}\,, \quad \delta_\epsilon\psi = \Lie_\epsilon\psi = \epsilon^\mu\partial_\mu\psi
\ee
where $\epsilon$ is any vector field. From the properties of the Lie derivative it follows that
\be
[\delta_\xi,\delta_\epsilon] = \delta_{[\xi,\epsilon]}
\ee
and the Lie algebra of the diffeomorphism group is given by the Lie algebra of vector fields.

In UD, the action of symmetry generators is different: we now need to act with a compensating Weyl transformation to ensure that $g$ remains unimodular. These ``Weyl-compensated'' diffeomorphisms are still generated by vector fields, but their infinitesimal action is
\be
\delta_\epsilon g_{\mu\nu} = \Lie_\epsilon g_{\mu\nu} - \frac{1}{2} g_{\mu\nu}\nabla_\rho\epsilon^\rho\,,\quad \delta_\epsilon\psi = \epsilon^\mu\partial_\mu\psi + \frac{1}{4}\psi\nabla_\rho\epsilon^\rho\,.
\label{newdiffs}
\ee
It is now easy to check explicitly that 
\be
[\delta_\xi,\delta_\epsilon]g_{\mu\nu} = \Lie_{[\xi,\epsilon]} g_{\mu\nu} - \frac{1}{2} g_{\mu\nu}\nabla_\rho[\xi,\epsilon]^\rho\,,\quad 
[\delta_\xi,\delta_\epsilon] \psi = [\xi,\epsilon]^\mu\partial_\mu\psi + \frac{1}{4}\psi\nabla_\rho[\xi,\epsilon]^\rho\,.
\ee
Hence we have, again, $[\delta_\xi,\delta_\epsilon] = \delta_{[\xi,\epsilon]}$, the Lie algebra of $\diff$. In order to clarify that this group acts differently on fields than usual diffeomorphisms, we call the group generated by (\ref{newdiffs}) $\diff^*$.
\smallskip

At this point we can close the circle of theories by noting that
the minimal formulation of UG can be obtained from UD
by fixing the scalar
\be
\psi=\psi_0\equiv\sqrt{\frac{Z_N}{\alpha}}\,.
\label{olga}
\ee
This fixes the Weyl invariance leaving just the group $\sdiff$.

\subsection{$\wtdiff$ gravity}

As a side remark, we observe that in the literature
one finds also another closely related action for GR
which can be obtained from DG as follows.
Start from the action (\ref{actionD2}) and integrate by parts
the second term so as to eliminate the Laplacian.
Then fix the scalar field to be
\be
\psi=\psi_0\left(\frac{|g|}{\omega^2}\right)^{-\frac{d-2}{4d}}\,.
\label{lena}
\ee
This leads to the following action\footnote{Since under $\sdiff$ the determinant
$|g|$ is a scalar, we have $\nabla_\mu|g|=\partial_\mu|g|$.}
\cite{Alvarez:2006uu,Blas:2011ac,Oda:2016psn,Bonifacio:2015rea}
\be
S_{X}(g)=Z_N\int \ud^dx\, |g|^\frac{1}{d}\,\omega^\frac{d-2}{d}
\left[R+\frac{(d-1)(d-2)}{4d^2}
\left(|g|^{-1}\nabla|g|
-2\omega^{-1}\nabla\omega\right)^2\right]\ .
\label{ester}
\ee
The presence of an unconventional power of the determinant
breaks the diffeomorphism invariance to invariance under diffeomorphisms that preserve $|g|$.
This action is invariant under $\sdiff\ltimes\mathit{Weyl}$,
hence the name $\wtdiff$ ($T\!\diff$ being synonymous to
$\sdiff$).

Invariance under Weyl transformations is not immediately obvious.
It becomes obvious once we note that the same action
can also be obtained from the Hilbert action (\ref{actionE}) 
by setting $S_X(g)=S_{{\rm EG}}(\bg)$, where
\be
\bg_{\mu\nu}=
\left(\frac{|g|}{\omega^2}\right)^{-1/d}
g_{\mu\nu}\ .
\label{gamma}
\ee
The metric $\bg_{\mu\nu}$ is manifestly invariant
under Weyl transformations of $g_{\mu\nu}$, with fixed $\omega$.
Also, note that the metric $\bg_{\mu\nu}$ is unimodular
by construction, so this is also viewed
as a form of unimodular gravity:
instead of removing the determinant by a constraint on the metric,
it is removed by making the action independent of it.

\begin{figure}
\begin{center}
\begin{picture}(300,180)
\put(140,170){{\bf DG}}
\put(105,157){$(g,\psi);\;\diff\ltimes\mathit{Weyl}$}
\put(100,150){\vector(-1,-1){40}}\put(190,150){\vector(1,-1){40}}\put(145,150){\vector(0,-1){40}}
\put(35,130){$\psi=\psi_0$}\put(212,130){$\sqrt{|g|}=\omega$}\put(104,130){$\psi=\psi_0\left(\frac{|g|}{\omega^2}\right)^{\frac{2-d}{4d}}$}
\put(230,93){{\bf UD}}\put(50,93){{\bf EG}}\put(130,93){$WT\!\diff$}
\put(210,80){$(g_U,\psi);\;\diff^*$}\put(30,80){$(g);\;\diff$}\put(100,80){$(g);\;\sdiff\ltimes\mathit{Weyl}$}
\put(60,70){\vector(1,-1){45}}\put(230,70){\vector(-1,-1){45}}\put(145,70){\vector(0,-1){40}}
\put(32,45){$\sqrt{|g|}=\omega$}\put(212,45){$\psi=\psi_0$}\put(98,45){$\sqrt{|g|}=\omega$}
\put(140,15){{\bf UG}}
\put(115,2){$(g_U);\;\sdiff$}
\end{picture}
\caption{Dynamical fields and symmetry groups for different formulations of vacuum GR; $g_U$ stands for a metric constrained to be unimodular.}
\label{figure}
\end{center}
\end{figure}
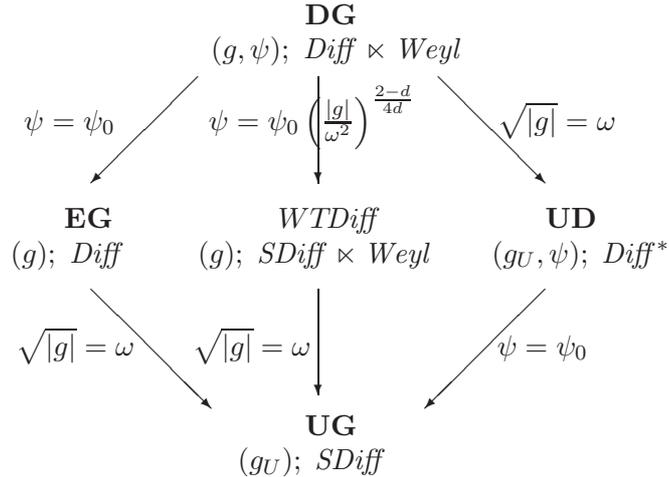

\subsection{Which description is preferable?}

The different formulations of GR discussed in the preceding 
subsections and their mutual relations are summarised in figure \ref{figure}. 
Given that they are all locally equivalent,
are there reasons to prefer one over another?
Actually there are good reasons to prefer the theories that
lie above in the figure 
and others to prefer the theories that lie below.

In gauge theories we deal with more variables than are strictly
necessary to describe the physical degrees of freedom of the system.
The main reason for accepting this redundancy and the complications
it entails is that in this way we can work with local fields\footnote{Another reason (for flat space QFT) is that this redundancy allows to preserve manifest
Lorentz invariance.}.
If we had good techniques for computing directly in terms of 
the physical, non-local variables, 
we would not need to deal with gauge invariances.
From this point of view, DG seems a perverse
complication of the description of GR.
In the literature, this point of view is represented
for example in \cite{jackiw,Oda:2016pok,Oda:2016psn},
where the Weyl invariance of DG is viewed as `fake'.
The characterisation of the Weyl invariance as fake 
was based on the vanishing of the Noether current.
We would like to propose here a different notion:
we will say that a gauge invariance is `inessential' 
if there is a way of fixing it which results in a 
description that is still in terms of local fields.
This is the case when there is a field that transforms
by a shift under infinitesimal gauge transformations.
The standard example is a nonlinear sigma model with values
in $G/H$ coupled to gauge fields for the group $G$.
One can partly gauge fix $G$ by fixing the nonlinear scalar,
leaving a massive YM field with values in the Lie algebra of $G$,
but only an $H$ gauge invariance.
By contrast, a simple example of an `essential'
gauge invariance is the one of QED.
Indeed, any attempt to fix the $U(1)$ gauge will result
in the new variables being related to the original ones
by a non-local transformation.
For example, if we fix the Lorentz gauge,
the remaining, physical, degree of freedom is a transverse
vector, which is obtained from the original gauge potential
acting with a projector that involves an inverse d'Alembertian.

For the reasons mentioned above, it seems desirable to 
eliminate all inessential gauge invariances and to work with the
smallest gauge group that is compatible with locality.
In the case of gravity, not only is the Weyl invariance of DG inessential,
but also part of the $\diff$ invariance of $EG$.
From this point of view, UG is the best formulation of GR\footnote{A related motivation for UG is the fact that a spin-two
particle is described by a {\it traceless} symmetric tensor
\cite{vanderBij:1981ym}, and so the trace is unnecessary.}.

However, there may be other arguments that may lead us
to accept a larger gauge group.
One such argument, of a rather formal nature,
appears from figure \ref{figure}.
EG, UD and $\wtdiff$-gravity all have the same number of gauge
generators (equal to the dimension of spacetime),
but the gauge group acts in different ways on the fields,
and may even be different.
The equivalence of these theories is apparent 
if we allow the gauge group to be enlarged,
until all three formulations are seen to be
different gauge fixings of a `linking theory', 
in the present case DG.
We refer to the transitions between EG, UD and $\wtdiff$
gravity as `symmetry trading', a notion which we will also discuss in more generality later on.

There is also another, more physical argument in
favour of larger gauge groups.
In gauge theories there is another notion of `locality',
different from the one discussed above:
the fields need not be defined continuously on all of spacetime.
It is enough that they can be defined on an open neighbourhood
of every point.
These locally defined fields may have singularities, but
what matters is that gauge invariant observables be regular.
Then if the field has a singularity somewhere, it must be possible
to find another, gauge-equivalent field that is is regular there.
Thus gauge fields are collections of locally defined fields,
such that the union of all their domains of definition
covers all of spacetime.
In Yang-Mills theory, the classic example that requires 
this generalised notion of field is the instanton.
In gravity, this notion is exemplified by the Schwarzschild horizon,
and by the coordinate transformation from Schwarzschild
to Kruskal coordinates.

There is a tension between these two notions of locality, 
in the sense that trying to minimise the
number of local fields in the description of a given system
may forbid the existence of certain solutions:
if the gauge group is smaller, a singularity in a local solution
is less likely to be a gauge singularity.
From this point of view, formulations with larger gauge group
could allow us to make sense of field configurations that would
otherwise seem to be singular. One example is the possible resolution of curvature singularities in the metric in formulations with a larger gauge group, as we shall discuss further in section \ref{discsec}.

In the core of this paper (sections \ref{HamiltonianDirac}-\ref{GfDirac})
we shall compare the Hamiltonian descriptions of DG, EG, UD, UG.
This Hamiltonian analysis would be the starting point for canonical quantization, which provides further arguments for or against different formulations of (locally) the same theory, and the choice of  different variables and symmetries. 

A prime example of such differences in canonical quantization is the introduction of Yang-Mills-type variables for GR in the Ashtekar formulation \cite{Ashtekar}, which allows a more rigorous construction of a Hilbert space for quantum GR when compared to the ADM formalism by using additional variables and a inessential gauge symmetry under local frame rotations. 

An example of a symmetry in the Hamiltonian formalism whose presence may obstruct canonical quantization is the refoliation invariance of all formulations we have discussed so far, expressed by the presence of a Hamiltonian constraint. This symmetry poses the infamous problem of time \cite{PoT}: no foliation (or choice of time function) is singled out over another, posing a serious obstacle to canonical quantization. Attempts to eliminate the refoliation symmetry of GR have led, {\em inter alia}, to the formalism of shape dynamics \cite{ShapeDyn,ShapeBook} which does not have refoliation invariance, is locally but not globally equivalent to GR, and thus fits naturally in our discussion of symmetry trading. 
Unlike theories such as DG or $\wtdiff$, however, shape dynamics is usually fundamentally formulated at the Hamiltonian rather than Lagrangian level. We shall discuss shape dynamics and its relation to EG in section~\ref{discsec}. An idea which we will be able to substantiate in some cases is that DG can provide a linking theory also for EG and shape dynamics, when expressed in Hamiltonian language.

\section{Hamiltonian analysis of DG}
\label{HamiltonianDirac}

We begin with the canonical formulation of DG. We will encounter curious properties of the Hamiltonian constraint in this formalism which, to our knowledge, have not been discussed in the literature before.  In this and the following sections we restrict 
our attention to $d=4$. 

DG can be seen as a limit of two better studied theories of gravity with a dynamical scalar: it corresponds to Brans-Dicke gravity \cite{bransdicke}
\bea
S_{{\rm BD}}(g,\phi)&=&\alpha\int \ud^4x\, \sqrt{|g|}\, \left[\phi^2 R
-4\omega g^{\mu\nu}\partial_\mu\phi\partial_\nu\phi\right]
\nonumber
\\&=&\alpha\int \ud^4x\, \sqrt{|g|}\, \left[\Phi R
-\frac{\omega}{\Phi} g^{\mu\nu}\partial_\mu\Phi\partial_\nu\Phi\right]\qquad (\Phi\equiv\phi^2)
\label{bransdicke}
\eea
where the Brans-Dicke parameter $\omega$ takes the value $\omega=-\frac{3}{2}$. This is the special value in which an additional (Weyl) symmetry emerges and the scalar field $\Phi$ no longer carries an independent physical degree of freedom. Brans-Dicke gravity with $\omega=-\frac{3}{2}$ is sometimes called `pathological' (see e.g.~\cite{Hammad:2018hhv}), for instance because the initial value problem in the sense of scalar-tensor theory \cite{Salgado:2005hx} is ill-posed. In the view we explore in this paper, the action (\ref{bransdicke}) with $\omega=-\frac{3}{2}$ is simply one way of defining GR; assuming that the coupling to matter preserves Weyl invariance, one has an additional gauge symmetry which is to be fixed in order to have a well-posed initial value problem. The theory, as it stands, is then no more `pathological' than EG.

Up to an overall sign,
DG is also the limit of the action for a scalar conformally coupled to GR,
\be
S_{{\rm GR+}\phi}(g,\phi)=Z_N\int \ud^4x\, \sqrt{|g|}\, R-\frac{1}{2}\int \ud^4x\, \sqrt{|g|}\, 
\left[g^{\mu\nu}\partial_\mu\phi\partial_\nu\phi+\frac{1}{6}\phi^2 R\right]\ ,
\ee
in which the Einstein-Hilbert term is absent ($Z_N\rightarrow 0$). The Hamiltonian analysis of Brans-Dicke gravity has been discussed in \cite{BDHamilt} and that of GR with a conformally coupled scalar in \cite{Kiefer,Bodendorfer}. It turns out that the limits $\omega\rightarrow-\frac{3}{2}$ or $Z_N\rightarrow 0$ in the Hamiltonian formulation of these theories are singular, as we will explain in more detail below, because of the additional Weyl symmetry that is present in the limit. This is why we present a self-consistent Hamiltonian analysis of DG in this section, which to our knowledge has not been discussed previously.

The starting action is (\ref{actionD2})
and we choose the parameter $\alpha$ to be given by (\ref{alpha}):
\be
S_{{\rm DG}}(g,\phi)=\frac{1}{12}\int \ud^4x\, \sqrt{|g|}\, 
\left[\phi^2 R
+6 g^{\mu\nu}\partial_\mu\phi\partial_\nu\phi\right]\ .
\label{actionGR4d}
\ee
We use the familiar ADM decomposition
\be
\ud s^2 = -N^2\ud t^2+q_{ij}(\ud x^i+N^i\ud t)(\ud x^j+N^j\ud t)
\label{ADMmetric}
\ee
and then express the time derivatives $\dot q_{ij}$
and $\dot \phi$ in terms of the extrinsic curvature 
\be
K_{ij}=\frac{1}{2N}\left(\dot q_{ij}
-D_iN_j-D_jN_i\right)
\label{ECurv2}
\ee
(where $D$ is the covariant derivative associated with the metric $q_{ij}$) and the normal derivative
\be
\partial_n \phi\equiv n^\mu\partial_\mu\phi=\frac{1}{N}\left(\dot\phi-N^j\partial_j\phi\right)
\ee
where $n^\mu$ is the normal vector to hypersurfaces of constant $t$.
\\In this way we arrive at (integrating by parts in the second step)
\bea
S_{{\rm DG}}(g,\phi)&=&\frac{1}{12}\int \ud t\int \ud^3x\, \sqrt{q}N\, 
\biggl[\phi^2 \left(R^{(3)}+K_{ij}K^{ij}-K^2\right)
\nonumber\\
&&
-6 (\partial_n \phi)^2
-4K\phi \partial_n \phi
+6 q^{ij}\partial_i\phi\partial_j\phi
+\frac{4}{N}\phi q^{ij}\partial_i\phi\partial_j N
\biggr]
\nonumber
\\&=&\frac{1}{12}\int \ud t\int \ud^3x\, \sqrt{q}N\, 
\biggl[\phi^2 \left(R^{(3)}+K_{ij}K^{ij}-K^2\right)
\nonumber\\
&&
-6 (\partial_n \phi)^2
-4K\phi \partial_n \phi
+2 q^{ij}\partial_i\phi\partial_j\phi
-4 \phi D^2\phi
\biggr]\,.
\label{actionadmd}
\eea
This form of the Lagrangian agrees with known expressions for GR with a conformally coupled scalar field \cite{Kiefer,Bodendorfer} in the appropriate limits ($Z_N\rightarrow 0,\;d=4$), which are well-behaved at the Lagrangian level. Boundary terms were neglected in \cite{Kiefer,Bodendorfer} and, as stated above, we shall not discuss them either.

\subsection{Constraints}
\label{Diracconst}

Denote $p^{ij}$ the momenta conjugate to $q_{ij}$,
$P$ and $P^i$ the momenta conjugate to $N$ and $N_i$
and $\pi$ the momentum conjugate to $\phi$.
We have the following relations between momenta
and velocities:
\bea
p^{ij}&=&\frac{1}{12}\sqrt{q}\phi^2
\left(K^{ij}-q^{ij}K-2q^{ij}\phi^{-1}\partial_n \phi\right)\,,
\label{defp}
\\
\pi&=&-\sqrt{q}\left(\partial_n \phi+\frac{1}{3}\phi K\right)\,.
\label{defpif}
\eea
There are no terms proportional to $\dot N$ and $\dot N^i$
in the action, so we have the usual primary constraints
\bea
P&=&0\ ,
\label{pc1}
\\
P_i&=&0\ .
\label{pc2}
\eea
In addition, the trace of (\ref{defp}),
$$
p\equiv q_{ij}p^{ij}=
-\frac{1}{2}\sqrt{q}\phi\left(\partial_n \phi+\frac{1}{3}\phi K\right)\ ,
$$
is proportional to (\ref{defpif}),
so we have another primary constraint
\be
\cC\equiv\phi\pi-2p=0\ .
\label{pc3}
\ee
Rewriting the Lagrangian in (\ref{actionadmd}) as $\mathcal{L}=p^{ij}\dot{q}_{ij}+\pi\dot\phi-H$, after a bit of work we obtain the primary Hamiltonian
\be
H_1\sim\cH[N]+\cD[\vec N]+\cC[\rho]
\label{hamiltonian}
\ee
where $\sim$ means `up to arbitrary terms proportional to the
primary constraints'
and $\cD$ and $\cH$ will be seen to correspond to the
diffeomorphism and Hamiltonian constraints. 
We have added a term with a Lagrange multiplier $\rho$
enforcing (\ref{pc3}) explicitly.
As usual, the constraints are smeared with suitable `test functions', here
\bea
\cC[\rho]&=&\int \ud^3x\; \rho(x)\cC(x)
\\
\cD[\vec N]&=&\int \ud^3x\;N^j(x)\cD_j(x)
\\
\cH[N]&=&\int \ud^3x\; N(x)\cH(x)
\eea
where
\be
\cD_j=\pi\partial_j\phi-2D_i p^i{}_j
\ee
and
\be
\cH(x)=
\frac{12}{\sqrt q\phi^2}\tilde p_{ij}\tilde p^{ij}
-\frac{1}{2\sqrt q}\pi^2
-\frac{1}{12}\sqrt q
\left(\phi^2 R^{(3)}+2q^{ij}\partial_i\phi\partial_j\phi
-4\phi D^2\phi\right)\ .
\label{Hamiltonian}
\ee
Here we denote $\tilde p^{ij}$ the tracefree part of the momentum 
\be
\tilde p^{ij}=p^{ij}-\frac{1}{3}q^{ij}p\ .
\ee
Let us repeat that at this point, due to (\ref{pc3}), the given expressions for $\cH$ and $\cD_i$ are only uniquely defined up to actions of $\cC$. We will aim to fix this ambiguity in the following discussion.

Preservation in time of the primary constraints
(\ref{pc1},\ref{pc2}) implies the secondary constraints
\bea
\cH&=&0\ ;
\label{sc1}
\\
\cD_i&=&0\ .
\label{sc2}
\eea
As we shall see below, there are no further constraints.

The dynamics of the lapse $N$ and shift $N^i$ and their conjugate momenta are rather trivial; the momenta are constrained to vanish whereas $N$ and $N^i$ are arbitrary spacetime functions to be chosen for convenience. The usual convention is then to remove $(N,P)$ and $(N^i,P_i)$ from the phase space and treat $N$ and $N^i$ as Lagrange multipliers enforcing the constraints. The Hamiltonian as given in (\ref{hamiltonian}) is then simply a sum of five first class constraints smeared with different Lagrange multipliers.

\subsection{Algebra of the constraints}
\label{constraintalgebra}

We now show that the constraints $\mathcal{H}$, $\mathcal{D}_i$ and $\mathcal{C}$ indeed form a closed (first-class) algebra\footnote{We ignore the constraints $P$ and $P_i$ from now on, as their Poisson brackets with all other constraints vanish.}. First, we illustrate their interpretation as generators of gauge transformations in DG.

The constraints $\cD_i$ generate infinitesimal diffeomorphisms.
Indeed, for any vector field $\epsilon^i$ we have
$$
\{X,\cD[\vec\epsilon]\}=\delta_\epsilon X=\cL_\epsilon X\ ,
$$
where $\cL_\epsilon$ denotes the Lie derivative along $\epsilon$.
For the canonical variables we have
\bea
\delta_\epsilon \phi&=&\epsilon^i\partial_i\phi\,,
\\
\delta_\epsilon\pi&=&\partial_i(\epsilon^i\pi)
=\epsilon^i D_i\pi+\pi D_i\epsilon^i\,,
\\
\delta_\epsilon q_{ij}&=&D_i\epsilon_j+D_j\epsilon_i\,,
\\
\delta_\epsilon p_{ij}&=&
\epsilon^k D_k p^{ij}
-p^{ik}D_k\epsilon^j
-p^{jk}D_k\epsilon^i
+p^{ij}D_k\epsilon^k
\,.
\eea
In the second and fourth relation one has to remember that
the momenta are densities of weight one,
and for subsequent manipulations it is more convenient to 
write the Lie derivatives in terms of covariant
derivatives instead of partial derivatives.
We also note the relation
\be
\delta_\epsilon\sqrt{q}=\partial_k(\epsilon^k\sqrt q)\ .
\ee

Similarly one finds that the constraint $\cC[\omega]$
generates infinitesimal Weyl transformation with 
parameter $\omega$:
$$
\{X,\cC[\omega]\}=\delta_\omega X\,,
$$
and more specifically for the canonical variables:
\bea
\delta_\omega\phi&=&\omega\phi\,,
\\
\delta_\omega\pi&=&-\omega\pi\,,
\\
\delta_\omega q_{ij}&=&-2\omega q_{ij}\,,
\\
\delta_\omega p^{ij}&=&2\omega p^{ij}\,.
\eea
We also note the relation
\be
\delta_\omega\sqrt{q}=-3\omega\sqrt q\,.
\ee
Using these relations one finds that the constraints
obey the following algebra:
\bea
\{\cD[\vec\epsilon_1],\cD[\vec\epsilon_2]\}
&=&\cD[[\vec\epsilon_1,\vec\epsilon_2]]\ ,
\\
\{\cC[\omega_1],\cC[\omega_2]\}
&=&0\ ,
\\
\{\cD[\vec\epsilon],\cC[\omega]\}
&=&\cC[\epsilon^i\partial_i\omega]\ ,
\\
\{\cD[\vec\epsilon],\cH[N]\}
&=&\cH[\epsilon^i\partial_i N]\ ,
\\
\{\cC[\omega],\cH[N]\}
&=&\cH[\omega N]\ .
\eea

Using these relations, we will see that imposing the time preservation
of the constraints does not generate new constraints,
and that the constraints form a first class system.

The only remaining bracket is that of two Hamiltonian constraints. In the ADM formulation of canonical GR, this bracket is equal to a spatial diffeomorphism that depends on phase-space functions; the four constraints $\cD_i$ and $\cH$ then satisfy the {\em hypersurface-deformation algebra} which is interpreted geometrically as the statement that these constraints generate four-dimensional spacetime diffeomorphisms. If we start from the symmetries and seek a spacetime-covariant theory of a metric and conjugate canonical momentum by imposing the presence of constraints satisfying the hypersurface-deformation algebra, the dynamics of GR is essentially unique \cite{GRregained}.

In the case of DG, this discussion becomes more subtle. In particular, with the definition (\ref{Hamiltonian}) we find 
\be
\{\cH[N],\cH[M]\} = \cD[q^{ij}(N\partial_iM-M\partial_iN)] -\frac{1}{3}\cC[N D^2 M - M D^2 N]\ .
\ee
The algebra thus remains first class, and no further constraints are generated, but the commutator of two gauge transformations generated by $\cH$ is not simply a spatial diffeomorphism: 
it involves a Weyl transformation. Since its action mixes diffeomorphisms with Weyl transformations, $\cH$ is not simply a generator of refoliations as in the ADM formalism.

To find an alternative definition of $\cH$ that is simply a generator of refoliations, we have the freedom of using the primary constraint $\cC$. In particular, we can remove the $\pi^2$ term from $\cH$ in exchange for changing the coefficient of $p^2$, i.e.
\be
\tilde{\mathcal{H}}(x)\equiv \frac{12}{\sqrt{q}\phi^2}\left(p_{ij}p^{ij}-\frac{1}{2}p^2\right)-\frac{\sqrt{q}}{12}\left(\phi^2 R^{(3)}+2q^{ij}\partial_i\phi\partial_j\phi-4\phi D^2\phi\right)
\label{newHamiltonian}
\ee
(this form will be used again in the following). This constraint satisfies
\be
\{\tilde{\mathcal{H}}[N],\tilde{\mathcal{H}}[M]\}=\cD[q^{ij}(N\partial_iM-M\partial_iN)]+\cC\left[q^{ij}(N\partial_iM-M\partial_iN)\phi^{-1}\partial_j\phi\right]
\label{commut2}
\ee
and thus has the same issue as far as its geometric interpretation is concerned.

The statement can be sharpened: consider now a general Hamiltonian constraint of the form
\be
c_1\frac{p_{ij}p^{ij}}{\sqrt{q}\phi^2}+c_2\frac{p^2}{\sqrt{q}\phi^2}+c_3\frac{\pi^2}{\sqrt q}+c_4\frac{p\pi}{\sqrt q\phi}+c_5\sqrt{q}\phi^2 R^{(3)}+c_6 \sqrt{q}q^{ij}\partial_i\phi\partial_j\phi+c_7\sqrt{q}\phi D^2\phi
\label{linearcomb}
\ee
where the $c_i$ are initially free coefficients. The seven terms we include are all scalar densities under spatial diffeomorphisms with the correct conformal weights, as it must be if we want to maintain the correct transformation behaviour under transformations generated by $\cD_i$ and $\cC$.

We then ask whether there is a choice of $c_i$ such that the resulting Hamiltonian constraint $\hat{\cH}$ defines the dynamics of DG (i.e.~is equivalent to (\ref{Hamiltonian}) up to the action of other constraints) and also satisfies the hypersurface-deformation algebra
\be
\{\hat{\cH}[N],\hat{\cH}[M]\} = \cD[q^{ij}(N\partial_iM-M\partial_iN)]\,.
\label{correctcommutator}
\ee
Intermediate steps of this calculation
are given in Appendix \ref{Appendix}. There is a somewhat trivial ambiguity related to a rescaling $\phi\rightarrow \lambda\phi$ and $\pi\rightarrow \lambda^{-1}\pi$ where $\lambda$ is a constant, which we fix by setting $c_5=-\frac{1}{12}$. We then find that the most general choice for the other $c_i$ in (\ref{linearcomb}) consistent with (\ref{correctcommutator}) leads to a Hamiltonian constraint
\be
\mathcal{H}_{{\rm BD}}\equiv \frac{12}{\sqrt{q}\phi^2}\left(p_{ij}p^{ij}-\frac{1}{2}p^2\right)-\frac{\sqrt{q}}{12}\left(\phi^2R^{(3)}+2q^{ij}\partial_i\phi \partial_j\phi-4\phi D^2\phi\right)-\frac{a}{4\sqrt{q}\phi^2}\mathcal{C}^2-\frac{1}{a}\sqrt{q}q^{ij}\partial_i\phi \partial_j\phi
\label{generalHam}
\ee
with only a single free parameter $a$. We have rearranged terms such that $\mathcal{H}_{{\rm BD}}=\tilde{\cH}+\cH^{(a)}$ where $\tilde{\cH}$ is a possible Hamiltonian constraint for DG and $\cH^{(a)}$ contains the two last terms that depend on the parameter $a$. The last term in (\ref{generalHam}) explicitly breaks the Weyl symmetry of the Hamiltonian constraint; Weyl symmetry, and the dynamics of DG, would only be recovered for $a\rightarrow\pm\infty$ which is not possible. We thus conclude that there is no Hamiltonian constraint for DG of the form (\ref{linearcomb}) that satisfies (\ref{correctcommutator}).

In fact, (\ref{generalHam}) is the Hamiltonian constraint of Brans-Dicke gravity (\ref{bransdicke}) with $a\equiv -\frac{6}{2\omega+3}$. Again, we see that $\omega\rightarrow-\frac{3}{2}$ in Brans-Dicke gravity would be $a\rightarrow\pm\infty$ in (\ref{generalHam}), so this limit cannot be taken to obtain a Hamiltonian constraint of DG. The same conclusion applies to a possible $Z_N\rightarrow\infty$ limit in the Hamiltonian formalism of GR conformally coupled to a scalar as presented in \cite{Bodendorfer}, where the Hamiltonian constraint also contains a term proportional to $\mathcal{C}^2$ whose coefficient diverges as $Z_N\rightarrow\infty$. In \cite{BDHamilt} the Hamiltonian constraint of Brans-Dicke gravity with $\omega=-\frac{3}{2}$, i.e.~of DG, 
was obtained by dropping the $\cC^2$ term; one then obtains $\tilde{\cH}$ in (\ref{newHamiltonian}) which, as we showed, does not satisfy the standard hypersurface-deformation algebra. 

Finally we note that there is a trick by which one can obtain (\ref{newHamiltonian}) directly from the (ADM) Hamiltonian constraint of EG \cite{kucharletter}. Namely, starting from 
\be
\mathcal{H}_{{\rm ADM}}=\frac{1}{Z_N\sqrt{\bar{q}}}\left(\bar{p}_{ij}\bar{p}^{ij}-\frac{1}{2}\bar{p}^2\right)-Z_N\sqrt{\bar{q}}R^{(3)}[\bar{q}]
\ee
for a 3-metric $\bar{q}_{ij}$ and conjugate momentum $\bar{p}^{ij}$, change variables as in (\ref{substitute}),
\be
\bar{q}_{ij}=\frac{1}{12 Z_N}\phi^2 q_{ij}\,,\quad 
\bar{p}^{ij}=12 Z_N\phi^{-2} p^{ij}\,.
\label{changevar}
\ee
Now use the formula for the transformation of the Ricci scalar under a conformal rescaling
and rescale the lapse by $\bar{N}=\sqrt{\frac{1}{12 Z_N}}\phi\,N$ to obtain the new Hamiltonian constraint that is exactly (\ref{newHamiltonian}). By construction, this method produces a Hamiltonian constraint that cannot depend on $\pi$, so it cannot resolve the ambiguities we have identified.

\section{Hamiltonian analysis of minimal UG}

Minimal UG is defined by the constraint (\ref{constraint}).
Following \cite{Unruh:1988in} we will impose this constraint 
at the Lagrangian level.
In the ADM decomposition it amounts to fixing the lapse to
\be
N=\frac{\omega}{\sqrt q}\ .
\label{admug}
\ee
We perform the 3+1 decomposition of the UG action (\ref{actionUG}) to obtain
\be
S_{{\rm UG}}(g)=Z_N\int \ud t\int \ud^3x\, \omega\, 
\left(K_{ij}K^{ij}-K^2+R^{(3)}\right)\,.
\label{actionadmug}
\ee
The Hamiltonian analysis is based on this Lagrangian.
It closely parallels the usual ADM formulation of EG, except that
there is no lapse variable.

The relation between $q_{ij}$ and their conjugate momenta $p^{ij}$
is given, as in EG, by
\be
p^{ij} =  Z_N\sqrt{q}\left(K^{ij}-q^{ij}K\right)\ ,
\ee
but now we have only the primary constraint (\ref{pc2}).

Proceeding as for DG, one finds that the primary Hamiltonian is
\be
H_1\sim\cH\left[\frac{\omega}{\sqrt{q}}\right]+\cD[\vec N]\ .
\ee
Consistency of the primary constraints under time evolution
leads to the secondary constraints (\ref{sc2}).
It is worth stressing that even though the gauge symmetry
at the Lagrangian level consists only of special diffeomorphisms,
the constraints (\ref{sc2}) generate the full 3-dimensional
diffeomorphism group, as in EG.
In this connection we note that given any spatial
vector field $\epsilon$,
it is always possible to extend it to a vector field on spacetime
that is divergence-free.
Thus the restriction on the gauge transformations is not
evident at this stage.

At this point the analysis deviates from that of EG.
Consistency of the secondary constraints implies that
\be
0=\left\{\cD[\epsilon],\cH\left[\frac{\omega}{\sqrt{q}}\right]\right\}
=\cH\left[D_i\left(\epsilon^i\frac{\omega}{\sqrt{q}}\right)\right]
\ ,
\ee
which implies the tertiary constraint
\be
D_i\cH=0\qquad
\mathrm{or}
\qquad
\cH=\sqrt{q}\,C\ ,
\label{tertiary}
\ee
for some integration constant $C$.
That is, the dynamical variables have to satisfy the ADM Hamiltonian constraint
with a cosmological term (where the value of $\Lambda$ is in general different for each solution)
\be
\cH_\Lambda=\frac{1}{Z_N\sqrt{q}}
\left(p_{ij}p^{ij}-\frac{1}{2}p^2\right)
-Z_N\sqrt{q}\left(R^{(3)}-2\Lambda\right)\ .
\ee
There are no further constraints.
Compared to the standard ADM formalism, there is 
one less first class constraint ($P=0$) but also one fewer canonical
pair of phase space variables ($N$, $P$).

The constraints are now identical to those of the ADM formulation of EG (including a cosmological constant $\Lambda$), and thus satisfy the hypersurface-deformation algebra
\bea
\{\cD[\vec\epsilon_1],\cD[\vec\epsilon_2]\}
&=&\cD[[\vec\epsilon_1,\vec\epsilon_2]]\ ,
\\
\{\cD[\vec\epsilon],\cH[N]\}
&=&\cH[\epsilon^i\partial_i N]\ \ ,
\\
\{\cH[N],\cH[M]\}
&=&\cD\left[q^{ij}(N\partial_i M-M\partial_i N)\right]\,.
\eea
They form a first class system.

\section{Hamiltonian analysis of UD}
\label{HamUDSec}

This proceeds very similarly to the case of minimal UG, with one small twist. As for minimal UG and EG, the only difference between UD and DG when deriving the Hamiltonian is that the lapse is fixed by the unimodularity condition, and we find a primary Hamiltonian
\be
H_1\sim\cH\left[\frac{\omega}{\sqrt{q}}\right]+\cD[\vec N]+\cC[\rho]\,.
\ee
The primary constraints are $\cC$ and $P^i$; there is again no primary constraint associated to the lapse, which does not appear in the action. Consistency of $P_i$ again leads to $\mathcal{D}_i$ as secondary constraints. 

However, we now have the additional constraint $\cC$ which also needs to be preserved under time evolution. This leads to the condition
\be
0=\left\{\cC[\rho],\cH\left[\frac{\omega}{\sqrt{q}}\right]\right\} = -2\cH\left[\frac{\rho\omega}{\sqrt{q}}\right]\ ,
\ee
i.e. it is now $\mathcal{H}$, not its derivative, which emerges as a tertiary constraint. In UD, unlike in UG, there is no additional global degree of freedom corresponding to the cosmological constant, as the Weyl symmetry forbids the appearance of a new dimensionful parameter (notice that $C$ in (\ref{tertiary}) has positive mass dimension 4).

\section{Gauge fixings of DG}
\label{GfDirac}

The way in which various theories descend from DG
by gauge fixing in the Lagrangian formalism
is summarised  in figure \ref{figure}.
We shall now discuss this at the Hamiltonian level for three cases of interest: the reduction from DG to EG and UD, 
and a gauge fixing of refoliation invariance.

\subsection{Einstein gauge}
\label{Einsteingauge}
We can recover the Hamiltonian form of EG by gauge fixing the Weyl symmetry of Hamiltonian DG. One adds a gauge fixing condition as an additional constraint,
\be
\mathcal{G}_E \equiv \phi - \phi_0 = 0
\ee
where $\phi_0$ is a non-zero constant with dimensions of mass. This gauge fixing corresponds to a global choice of units: we can identify $\phi_0=:\pm\sqrt{12Z_N}$ as corresponding to Newton's constant. The new gauge-fixing constraint is second class with respect to $\mathcal{C}$,
\be
\{\mathcal{C}[\rho],\mathcal{G}_E[\lambda]\} = -\int \ud^3 x\;\rho\lambda\phi\,.
\ee
That is, preservation of $\mathcal{G}_E$ under time evolution does not generate an additional constraint, but instead fixes the Lagrange multiplier $\rho$. We also have
\bea
\{\mathcal{D}[\vec\epsilon],\mathcal{G}_E[\lambda]\}&=& \int \ud^3 x\;\epsilon^i\,\partial_i\phi \approx 0\,,
\label{weakeq}
\\\{\tilde{\mathcal{H}}[N],\mathcal{G}_E[\lambda]\} &=& 0
\eea
where $\tilde{\mathcal{H}}$ is the redefined Hamiltonian constraint (\ref{newHamiltonian}), which is equivalent to $\mathcal{H}$ up to actions of the primary constraint $\mathcal{C}$. The notation $\approx$ in (\ref{weakeq}) is Dirac's notion of weak equality, i.e. equality when the constraints are satisfied.

The second class constraints $\mathcal{G}_E$ and $\mathcal{C}$ can now be solved for $\phi$ and $\pi$,
\be
\phi=\phi_0\,,\quad \pi= 2p\,\phi_0^{-1}\,,
\label{gaugefix1}
\ee
and $\phi$ and $\pi$ can be removed from the phase space (this is equivalent to the Dirac bracket procedure). We are left with the ADM phase space and constraints
\bea
\mathcal{D}_i\big|_{\phi^2= 12Z_N} & = & -2D_k{p^k}_i\,,
\\\mathcal{H} = \tilde{\mathcal{H}}\big|_{\phi^2= 12Z_N} &=& \frac{1}{Z_N\sqrt{q}}\left(p_{ij}p^{ij}-\frac{1}{2}p^2\right)-Z_N\sqrt{q}R^{(3)}\,,
\eea
and hence recover Hamiltonian EG.

For Einstein gauge to be available, it is necessary and sufficient that the solution of DG one considers does not have a zero of $\phi$. The breakdown of this gauge condition is associated with certain singularities in EG, as we will discuss below.

\subsection{Unimodular gauge}

The unimodular condition $\sqrt{|g|}=\omega$ can be used to fix the lapse, as we have discussed. In order for this to be a gauge fixing in the usual Hamiltonian sense, we need to return to a larger phase space for DG which contains the lapse $N$ and its momentum $P$ 
(we could also include the shift $N^i$ and its momentum $P_i$, but doing this would have no effect on what follows).

Let us then define
\be
\mathcal{P}[\tau]=\int \ud^3 x\; P(x)\tau(x)
\ee
as the smeared constraint associated to $P$, and add the gauge-fixing constraint
\be
\mathcal{G}_U \equiv N\sqrt{q} - \omega = 0\,.
\ee
This is second class with $\mathcal{P}$, since
\be
\{\mathcal{P}[\tau],\mathcal{G}_U[\lambda]\} = -\int \ud^3 x\,\sqrt{q}\,\lambda\tau
\ee
is not an additional constraint but fixes $\tau=0$. For the other constraints we find
\bea
\{\mathcal{H}[\epsilon],\mathcal{G}_U[\lambda]\} & = & 0\,,
\\\{\mathcal{D}[\vec\epsilon],\mathcal{G}_U[\lambda]\} & = & -\int \ud^3 x\,\sqrt{q}\,N\lambda (D_i\epsilon^i)\,,
\\\{\mathcal{C}[\rho],\mathcal{G}_U[\lambda]\} &=& 3\int \ud^3 x\,\sqrt{q}\,N\lambda \rho\,.
\eea
$\mathcal{H}[\epsilon]$ and the combinations $\mathcal{D}[\vec\epsilon]-\mathcal{P}[N\,D_i\epsilon^i]$ and $\mathcal{C}[\rho]+\mathcal{P}[3N\rho]$ are first class with the unimodular gauge fixing. We can then remove $N$ and $P$ from the phase space by solving for $N=\omega/\sqrt{q}$ and $P=0$. 
The resulting canonical formalism is just the one for UD 
that we discussed in section \ref{HamUDSec}: adding $\mathcal{G}_U$ is equivalent to choosing a specific form of the lapse.

We do not discuss in detail the other two arrows in Figure 1,
namely those from EG to UG and from UD to UG,
both of which amount to imposing the same gauge conditions
as in this and in the preceding sections.
Instead, we consider a different type of gauge-fixing
that has other virtues.

\subsection{Lichnerowicz gauge}

Here one fixes the refoliation invariance of GR, i.e.~gauge transformations generated by the Hamiltonian constraint $\mathcal{H}$. One simple gauge-fixing condition that can achieve this is
\be
\mathcal{G}_L \equiv \pi = 0\,. 
\ee
Through the conformal constraint $\mathcal{C}$, this condition is equivalent to imposing {\em maximal slicing} $p=0$. The Lichnerowicz gauge fixing is first class with $\mathcal{C}$ and $\mathcal{D}$,
\bea
\{\mathcal{D}[\vec\epsilon],\mathcal{G}_L[\lambda]\}&=& \mathcal{G}_L[\epsilon^i\partial_i\lambda]\,,
\\\{\mathcal{C}[\rho],\mathcal{G}_L[\lambda]\} &=& \mathcal{G}_L[\rho\lambda]\,.
\eea
On the other hand, the bracket of $\mathcal{G}_L$ with $\mathcal{H}$ leads to a {\em lapse-fixing equation} 
\be
\{\mathcal{H}[N],\mathcal{G}_L[\lambda]\} = \int \ud^3 x\;\lambda\left[-\frac{24 N}{\sqrt{q}\phi^3}\tilde{p}^{ij}\tilde{p}_{ij} - \frac{N}{6}\sqrt{q}\phi R^{(3)}+\sqrt{q}q^{ij}\partial_i N \partial_j \phi + \sqrt{q}N\,D^2\phi + \frac{\sqrt{q}}{3}\phi\,D^2N\right]
\ee
which is again an equation fixing a Lagrange multiplier and not an additional constraint: $\mathcal{G}_L$ is second class with $\mathcal{H}$ (notice that we have now returned to the conventional formalism in which $N$ is not part of the phase space but a Lagrange multiplier). We can use the Hamiltonian constraint and $\mathcal{G}_L$ to eliminate the $\tilde{p}_{ij} \tilde{p}^{ij}$ term from the lapse-fixing equation, reducing it to 
\be
\left(- \phi R^{(3)}-\phi^{-1}q^{ij}\partial_i \phi \partial_j \phi +5 D^2\phi\right) N+3q^{ij}\partial_i N \partial_j \phi+\phi\,D^2N = 0
\ee
or, somewhat more succinctly (see \cite{Anderson:2004wr}),
\be
\left(\chi^2 R^{(3)}-7 \chi D^2\chi\right) N - \chi^{-1} D^2(\chi^3 N)=0
\ee
with $\chi\equiv\phi^{1/2}$.
Again, the second class constraints $\mathcal{G}_L$ and $\mathcal{H}$ can now be solved for $\phi$ and $\pi$ to remove these variables from the phase space. The Hamiltonian constraint becomes, using $\pi=0$,
\be
\frac{12}{\sqrt{q}\phi^2}\tilde{p}_{ij}\tilde{p}^{ij}-\frac{\sqrt{q}}{12}\left(\phi^2 R^{(3)}+2q^{ij}\partial_i\phi\partial_j\phi-4\phi D^2\phi\right)=0\,.
\ee
This is the {\em Lichnerowicz equation} for $\phi$, which provides one way of finding suitable initial data for canonical GR on a spatial hypersurface. Again, its form simplifies when expressed in terms of $\chi\equiv\phi^{1/2}$,
\be
\frac{144}{q\,\chi^7}\tilde{p}_{ij}\tilde{p}^{ij}-\chi R^{(3)}+8 D^2\chi=0\,.
\ee
The remaining constraints are then simply
\be
\mathcal{C}\big|_{\pi=0} = -2p\,,\quad \mathcal{D}_i\big|_{\pi=0} = -2 D_k {p^k}_i\,.
\ee
That is, $p_{ij}$ must be transverse and traceless. 
\\
\\The meaning of the Lichnerowicz equation for the initial value problem of GR is the following: start with an arbitrary 3-metric $\bar{q}_{ij}$ and a symmetric tensor $\bar{K}^{ij}$ which is transverse and traceless with respect to $\bar{q}_{ij}$. Solve the Lichnerowicz equation for $\chi$; then $g_{ij}=\chi^4\,\bar{q}_{ij}$ and $K^{ij}=\chi^{-10}\,\bar{K}^{ij}$ provide initial data in terms of a 3-metric and extrinsic curvature, solving the Einstein equations. (See e.g.~\cite{ShapeBook} for more details and the extension of Lichnerowicz's to York's method.)
\\
\\In DG, the combinations $\phi^2\,q_{ij}$ and $\phi^{-5}\,\bar{K}^{ij} \equiv \phi^{-5}\,p^{ij}/\sqrt{q}$ are Weyl invariants; they hence reduce to the corresponding GR solutions in a gauge where $\phi$ is constant.
\\
\\The limitation of Lichnerowicz gauge is that it requires the existence of a global foliation for which $p=0$ everywhere. On compact manifolds, this is only possible for Yamabe-positive metrics which is a serious restriction. It would be desirable to extend Lichnerowicz gauge to a more general `York gauge' in which $p=\sqrt{q}\,\tau$ where $\tau$ is constant on each spatial slice but can differ from one slice to the next. This condition breaks conformal symmetry if $\tau$ is a scalar under Weyl transformations.

\section{Discussion}
\label{discsec}

\subsubsection*{Count of the degrees of freedom}

To summarise the results of the Hamiltonian analyses,
we compare here the count of degrees of freedom
in the standard formulation of GR (EG),
in the formulation with a dilaton (DG) 
and in unimodular formulation (UG), 
returning to a general number $d\ge 3$ of spacetime dimensions.

The count is summarised in table \ref{Table},
which reads as follows.
In the standard Einstein formulation there are 
$d(d+1)$ canonical variables and $2d$
first class constraints.
Since each first class constraint has to be accompanied
by a gauge condition, the total number of canonical degrees of freedom
is $d(d+1)-2\times2d=d(d-3)$.
In $d=4$ this agrees with the two polarization
states of the graviton.

In DG one has two more canonical variables
(the scalar field and its momentum).
There is also one more primary constraint and the same
number of secondary constraints, so the number of constraints
is one higher. This, and the associated gauge condition,
removes the additional variables.

In unimodular gravity there are two less canonical variables,
due to the condition of unimodularity of the four-metric,
which we use to eliminate the lapse
and the associated momentum.
There is then one less primary constraint than in
Einstein gravity, because the momentum conjugate to the lapse is
not a canonical variable.
There is also one less secondary constraint,
but then there is a tertiary constraint
(in another terminology, one would say that there
is the same number of secondary constraints).
The fact that there is one less constraint is related 
to the fact that the gauge group $\sdiff$
has one less free parameter.
Altogether, the constraints and their gauge condition
remove two variables less than in Einstein's theory,
so the final number of degrees of freedom is the same.

\begin{table}[htp]
\begin{center}
\begin{tabular}{ | l | l | l | l |}
\hline
  & {\bf DG} & {\bf EG} & {\bf UG} \\ \hline
Fields & $q_{ij}$,$N_i$,$N$,$\phi$
   & $q_{ij}$,$N_i$,$N$ 
   & $q_{ij}$,$N_i$\\ \hline
Momenta & $p^{ij}$,$P^i$,$P$,$\pi$ 
   & $p^{ij}$,$P^i$,$P$ 
   & $p^{ij}$,$P^i$ \\ \hline
\# of canonical variables & $d(d+1)+2$ 
   & $d(d+1)$ 
   & $d(d+1)-2$ \\ \hline
Primary constraints & $P^i$, $P$, $\cC$ 
   & $P^i$, $P$
   & $P^i$\\ \hline

Secondary constraints & $\cH_i$,$\cH$ & $\cH_i$,$\cH$ & $\cH_i$,$\cH_\Lambda$ \\ \hline
\# of first class constraints & $2d+1$ & $2d$ & $2d-1$  \\ \hline
\# of canonical d.o.f. & $d(d-3)$ 
& $d(d-3)$ 
& $d(d-3)$  \\ \hline
\end{tabular}
\end{center}
\caption{Summary of the constraint analysis of Dirac, Einstein and unimodular gravity.}
\label{Table}
\end{table}

\subsubsection*{Symmetry trading and linking theories}

Choosing Lichnerowicz gauge in DG leads to a theory that is different from EG: compared to EG, the refoliation invariance generated by $\mathcal{H}$ has been `traded' for a Weyl symmetry generated by $\cC$. The canonical variables are the same in both theories (a spatial metric and its conjugate momentum); the same number of first class constraints implies the same number of (local) degrees of freedom. One of the main points we want to convey in this paper is that the relation between two locally equivalent theories, sharing the same dynamical variables and the same number of gauge symmetries per spacetime point, is often easiest within a `linking theory', here given by DG.

In section \ref{introsec} we already saw an example of this in the covariant Lagrangian formalism (see figure \ref{figure}). Compare EG and $WT\!\diff$ gravity: they are both formulations of GR in terms of a Lorentzian metric, but with different action and different symmetries. Understanding the relation of the two can be done either via UG \cite{Alvarez:2016lbz} or, as we have done here, via DG. As we have seen, EG and $WT\!\diff$ are obtained by different choices for the scalar $\psi$ (or $\phi$, the two coincide in $d=4$) in DG. Thus, DG provides a `linking theory' from which EG and $WT\!\diff$  can be obtained rather straightforwardly. A linking theory, when it exists, also clarifies the global differences between different formulations, which correspond to the failure of the respective gauge-fixing conditions for certain solutions, as we will discuss 
in the section `Singularities' below. Symmetry trading then becomes more directly understandable in terms of the linking theory.

\subsubsection*{Shape dynamics}

DG in Lichnerowicz gauge provides a dynamical system for gravity that is similar to shape dynamics \cite{ShapeDyn,ShapeBook}, the key difference being that Weyl transformations in DG are not constrained to be volume-preserving; accordingly, unlike shape dynamics, DG does not have a global Hamiltonian\footnote{We have dropped boundary terms in our analysis, e.g.~by assuming that the spacetime manifold is compact without boundary. Boundary terms would lead to a true Hamiltonian, corresponding to the ADM energy.}.
It is known that both Hamiltonian (ADM) EG and shape dynamics can be seen as gauge fixings of a Hamiltonian linking theory \cite{ShapeDyn}. 
One of the original motivations for the present work 
was to seek a spacetime covariant, Lagrangian form for the linking theory. 
DG provides the obvious candidate for relating 
Hamiltonian EG to shape dynamics. 

Under certain further assumptions, the known linking theory for EG and shape dynamics can indeed be seen to be equivalent to DG. The following system of constraints for a 3-metric $q_{ij}$ and canonical momentum $p^{ij}$, scalar field $\Phi$ and momentum $\pi_\Phi$ provides a linking theory for shape dynamics \cite{ShapeBook}:
\bea
\mathcal{H}&=& \frac{e^{-6\hat\Phi}}{\sqrt{q}}\tilde{p}^{ij}\tilde{p}_{ij}-\frac{e^{-6\hat\Phi}}{6\sqrt{q}}\left(\frac{\pi_\Phi}{4}+e^{6\hat\Phi}\sqrt{q}\langle p\rangle\right)^2 -\sqrt{q}\left( R^{(3)} e^{2\hat\Phi}-8e^{\hat\Phi}\,D^2 e^{\hat\Phi}\right)\,,
\label{linking1}
\\\mathcal{H}_i &=& -2D_j{{p_i}^j}+\pi_\Phi\partial_i\Phi\,,
\\\mathcal{Q}&=&\pi_\Phi-4(p-\langle p \rangle\sqrt{q})\,.
\label{linking3}
\eea
Here $\tilde{p}_{ij}$ is again the tracefree part of $p_{ij}$, and $\hat\Phi\equiv\Phi-\frac{1}{6}\log\langle \sqrt{q}e^{6\Phi}\rangle$. Spatial averages for a scalar density $X$ are defined by $\langle X\rangle \equiv (\int \ud^3 x\,X)/(\int \ud^3 y\,\sqrt{q})$.

An elementary canonical transformation (for simplicity here, unlike in the rest of the paper, $\phi$ and $\Phi$ are treated as dimensionless)
\bea
e^{2\Phi}\equiv\phi\,,\quad \pi_\Phi = 2\pi\phi\ ,
\eea
transforms (\ref{linking1})-(\ref{linking3}) into
\bea
\mathcal{H}&=& \frac{\langle \sqrt{q}\phi^3 \rangle}{\sqrt{q}\phi^3}\tilde{p}^{ij}\tilde{p}_{ij}-\frac{\langle \sqrt{q}\phi^3 \rangle}{6\sqrt{q}\phi^3}\left(\frac{\pi \phi}{2}+\frac{\sqrt{q}\phi^3\langle p\rangle}{\langle \sqrt{q}\phi^3 \rangle}\right)^2 -\sqrt{q}\, \frac{\phi R^{(3)}-8\phi^{1/2}\,D^2 \phi^{1/2}}{\langle \sqrt{q}\phi^3 \rangle^{1/3}}\,,
\\\mathcal{H}_i &=& -2D_j{{p_i}^j}+\pi\partial_i\phi\,,
\\\mathcal{Q}&=&2\pi\phi-4(p-\langle p \rangle\sqrt{q})\,.
\eea
These constraints agree with the ones of DG (see section~\ref{Diracconst}) if
\be
\langle p\rangle = 0\,,\quad \langle \sqrt{q}\phi^3 \rangle={\rm const}.
\ee
At least in the specific case of configurations with vanishing average extrinsic curvature and with constant spatial volume as measured by the `Einstein frame' metric $\bar{q}_{ij}\propto \phi^2 q_{ij}$, 
DG is precisely the linking theory that was constructed for shape dynamics already. Its Hamiltonian formalism can be derived from a simple, generally covariant action (\ref{actionD2}) (with $d=4$).

\subsubsection*{Singularities}
\label{discsecsing}

The dilaton $\phi$ is naturally required to be positive
(in fact it is common to parametrize it as $\phi=f e^\sigma$).
However, in solving the equations one sometimes encounters
situations where it may be advantageous to relax such
a condition.

As an example, consider the case of a cosmological Friedmann-Lema\^itre-Robertson-Walker universe, which in EG can be written in terms of conformal time as
\be
\ud s^2 = a^2(\eta)(-\ud \eta^2+h_{ij}\ud x^i\ud x^j) \equiv a^2(\eta)\,g_{\mu\nu}^0 \ud x^\mu \ud x^\nu
\ee
where $h_{ij}$ is a fixed 3-metric of constant curvature. By (\ref{substitute}), it is now evident that such EG solutions can be `lifted' to DG solutions for which
\be
g_{\mu\nu} = g^0_{\mu\nu}\,,\quad \phi \equiv a(\eta) \sqrt{Z_N/\alpha}\,,
\ee
i.e. solutions with static spacetime metric and time-dependent $\phi$ field; $a(\eta)$ solves the Friedmann equations of usual cosmology.
From a mathematical point of view, the singularity has 
been shifted from the metric $g_{\mu\nu}$ to a zero of the dilaton. 
(Notice that $\phi\rightarrow 0$ means a divergence in the effective Newton's constant $\sim\phi^{-2}$.)
Whether this is to be regarded as a physical singularity
of the geometry depends on whether free falling test particles
are assumed to follow the geodesics of the metric $g_{\mu\nu}$
(in which case the physical singularity has been removed)
or of the original metric $\bar g_{\mu\nu}$ of (\ref{substitute})
(in which case it is still present).
In the former case, such cosmological solutions of DG allow an extension of the spacetime manifold through what would normally be the Big Bang/Big Crunch singularity at $a(\eta)=0$, the point where Einstein gauge breaks down. See \cite{DiracCosmo} for various recent works using DG and related formulations in this context in order to resolve the Big Bang/Big Crunch singularity of EG either classically or quantum mechanically. This discussion extends straightforwardly to general solutions of EG that have points or regions where the metric has vanishing determinant.

DG is most naturally coupled to matter fields that are themselves conformal (in the sense of being invariant under (\ref{fakeweyl})), and do not `see' any singularities in the conformal factor of the metric, nor in the dilaton $\phi$. For non-conformal matter there is, as usual, a choice of whether it couples only to the metric $g_{\mu\nu}$ or also to $\phi$, and coupling to $\phi$ would in general still lead to a singularity even if $g_{\mu\nu}$ is made to be non-singular.

Just as the larger gauge group of DG means that certain
singular fields in EG are not singular in DG,
the smaller gauge group of UG means that certain
regular fields in EG would be singular in UG.
As a simple example, in a 2-dimensional space with coordinates 
$x$ and $y$, consider $\omega=x$.
This is singular at $x=0$, because a volume form
is supposed to be nowhere-vanishing.
However, this is normally seen as a shortcoming of the
coordinates. Recognising that $\omega$ is the volume form
of flat space in polar coordinates, we remove the singularity
by simply passing to Cartesian coordinates.
However, in UG such a coordinate transformation would be
forbidden, so the singularity cannot be removed.
For a slightly less trivial example,
consider the Schwarzschild metric in isotropic coordinates
$$
\ud s^2 = -\left(\frac{r-M/2}{r+M/2}\right)^2 \ud t^2
+\left(1+\frac{M}{2r}\right)^4
(\ud x_1^1+\ud x_2^2+\ud x_3^2)\,.
$$
If we base UG on the fixed volume element
$\omega=\left(1-\frac{M}{2r}\right)\left(1+\frac{M}{2r}\right)^5$,
there is a (mild) 
singularity at $r=M/2$
that cannot be removed by gauge transformations.

It would be interesting to find other examples
of such behaviour in realistic solutions.

\subsubsection*{Cosmological constant term}

It would be straightforward to add a cosmological constant term to the actions we discussed. In DG in $d=4$, it would be of the form
\be
S_\Lambda=Z_N\int \ud^4 x\sqrt{|\bg|}\,2\Lambda\Big|_{\bg_{\mu\nu} = \frac{\alpha}{Z_N}\phi^2 g_{\mu\nu} }  
= \frac{\alpha^2}{Z_N}2\Lambda\int \ud^4 x\sqrt{|g|}\;\phi^4
\ee
i.e.~a $\phi^4$ potential with dimensionless coefficient, which is clearly allowed by Weyl invariance. Since such a term does not involve spatial derivatives, it does not affect the discussion of the constraint algebra in section \ref{constraintalgebra}. In UG, even with a cosmological constant term in the action one gets a different effective $\Lambda$ as an integration constant for each solution. We have omitted details of these discussions which do not seem important for the main points of our paper.

\subsubsection*{Further extensions}

As we hope to have made clear, the addition of inessential gauge invariances
can be useful in order to compare equivalent formulations of a theory
possessing different gauge invariances (symmetry trading).
We have discussed in some detail the case of Weyl invariance,
but one need not stop here.
There is a formulation of gravity where the gauge group is
extended by local $GL(4)$ transformations,
see \cite{Floreanini:1989hq} for a canonical analysis
and \cite{Percacci:2009ij} for a review and additional references. 
This formulation acts as a linking theory between EG
and the tetrad formulation, in the sense that both can
be obtained by fixing the value of some field\footnote{Incidentally, note that local Lorentz transformations
are inessential.}.
DG can be viewed as a sub-case of the $GL(4)$-formulation
where the $GL(4)$ transformations are restricted to multiples of
the identity.

We also note any metric can be locally transformed into any other
metric by means of a local $GL(4)$ transformation
$$
g'_{\mu\nu}(x)=\Lambda(x)_\mu{}^\alpha\Lambda(x)_\nu{}^\beta g_{\alpha\beta}(x)\ .
$$
A field theory of matter that is invariant
under such local $GL(4)$ transformations is
a theory that is independent of the metric 
and therefore is a topological field theory.
Just like a conformally invariant theory of matter
is insensitive to the cosmological singularity,
a $GL(4)$-invariant field theory will be insensitive
to {\it any} singularity.

\subsubsection*{Quantization}

Our main aim was to discuss inessential gauge invariance and
symmetry trading at the classical level, 
but we add here a few comments about these issues 
in a quantum context.
Weyl invariance is generally known 
to be anomalous at a quantum level \cite{duff}.
However, it has been known for a long time
that in the presence of a scalar field transforming
by a shift (a dilaton), it is possible to quantize the theory
preserving Weyl invariance
\cite{englert,fv,flop,reuterliouville,creh,shapo,Percacci:2011uf}.
This is therefore the case in DG.
For $\wtdiff$-gravity, this has been discussed in 
\cite{Alvarez:2013fs}.
It has also been argued on general grounds
\cite{Padilla:2014yea}
and shown explicitly in one-loop calculations
\cite{Gonzalez-Martin:2017fwz,Ardon:2017atk} that UG is equivalent to EG at the quantum level.
All this points to the conclusion that the local equivalence between
theories that differ by inessential gauge invariances
or symmetry trading can be maintained also at the quantum level.

\section*{Acknowledgments}

We would like to thank Julian Barbour and 
Flavio Mercati for stimulating discussions
on shape dynamics and for sharing with us reference \cite{kucharletter},
and the organisers of the `Shape Dynamics Workshop' that took place at the Perimeter Institute in May 2017, where this work was conceived.
This research was supported in part by Perimeter Institute for Theoretical Physics. Research at Perimeter Institute is supported by the Government of Canada through the Department of Innovation, Science and Economic Development Canada and by the Province of Ontario through the Ministry of Research, Innovation and Science. 

The work of SG was supported by a Royal Society University Research Fellowship (UF160622).

\appendix
\section{Evaluation of the Poisson bracket $\{\cH[N],\cH[M]\}$}
\label{Appendix}

For convenience of the reader, we summarise some details of how one evaluates the Poisson bracket $\{\cH[N],\cH[M]\}$ in DG and possible modifications (including Brans-Dicke gravity) as discussed above. We consider general Hamiltonian constraints that are a linear sum of the following terms:
\be
\frac{p_{ij}p^{ij}}{\sqrt{q}\phi^2}\,,\quad \frac{p^2}{\sqrt{q}\phi^2}\,,\quad \frac{\pi^2}{\sqrt q}\,, \quad \frac{p\pi}{\sqrt q\phi}\,, \quad \sqrt{q}\phi^2 R^{(3)}\,, \quad \sqrt{q}q^{ij}\partial_i\phi\partial_j\phi\,, \quad \sqrt{q}\phi D^2\phi
\ee
smeared with test functions $N(x)$ and $M(x)$.

The following nontrivial Poisson brackets appear in the required calculations:
\bea
\left\{\int \ud^3 x \frac{N\pi^2}{\sqrt{q}},\int \ud^3 y M\sqrt{q}\,q^{ij}\partial_i\phi\partial_j\phi\right\}-(\leftrightarrow)&=&4\int [N,M]^i\pi\partial_i\phi\,,
\\\left\{\int \ud^3 x \frac{N\pi^2}{\sqrt{q}},\int \ud^3 y M\sqrt{q}\,\phi D^2\phi \right\}-(\leftrightarrow)&=&\int [N,M]^i(-4\pi \partial_i\phi+2D_i(\pi\phi))\,,
\\\left\{\int \ud^3 x N\frac{p_{ij}p^{ij}}{\sqrt{q}\phi^2},\int \ud^3 y M\sqrt{q}\phi D^2\phi \right\}-(\leftrightarrow)&=&\int [N,M]^i\,\phi^{-1}\left(p\partial_i\phi-2{p_{i}}^j\partial_j\phi\right)\,,
\\\left\{\int \ud^3 x N\frac{p_{ij}p^{ij}}{\sqrt{q}\phi^2},\int \ud^3 y M\sqrt{q}R^{(3)}\phi^2 \right\}-(\leftrightarrow)&=&\int [N,M]^i\left(8p\phi^{-1}\partial_i\phi-2D_jp\right.\nonumber
\\&&\left.-8\phi^{-1}{p_i}^j\partial_j\phi+2D_j{p_i}^j\right)\,,
\\\left\{\int \ud^3 x N\frac{p^2}{\sqrt{q}\phi^2},\int \ud^3 y M\sqrt{q}\phi D^2\phi \right\}-(\leftrightarrow)&=&\int[N,M]^i p\phi^{-1}\partial_i\phi \,,
\\\left\{\int \ud^3 x N\frac{p^2}{\sqrt{q}\phi^2},\int \ud^3 y M\sqrt{q}R^{(3)}\phi^2 \right\}-(\leftrightarrow)&=&\int [N,M]^i\left(16p\phi^{-1}\partial_i\phi-4D_ip\right)\,,
\\\left\{\int \ud^3 x N\frac{p\pi}{\sqrt{q}\phi},\int \ud^3 y M\sqrt{q}q^{ij}\partial_i\phi\partial_j\phi\right\}-(\leftrightarrow)&=&2\int [N,M]^i p\phi^{-1}\partial_i\phi\,,
\\\left\{\int \ud^3 x N\frac{p\pi}{\sqrt{q}\phi},\int \ud^3 y M\sqrt{q}\phi D^2\phi \right\}-(\leftrightarrow)&=&\int [N,M]^i\left(D_ip-2p\frac{\partial_i\phi}{\phi}+\frac{\pi\partial_i\phi}{2}\right)\,,
\\\left\{\int \ud^3 x N\frac{p\pi}{\sqrt{q}\phi},\int \ud^3 y M\sqrt{q}R^{(3)}\phi^2 \right\}-(\leftrightarrow)&=&\int[N,M]^i\left(8\pi\partial_i\phi-2D_i(\pi\phi)\right)\,.
\eea
In these expressions, ``$(\leftrightarrow)$'' is a shorthand for taking the previous expression and exchanging $N$ and $M$, and we define $\int [N,M]^i V_i\equiv \int \ud^3 x\,q^{ij}(N\partial_jM-M\partial_jN)V_i$ as the right-hand side is always of this form. Where these expressions can be compared with those given in \cite{Bodendorfer}, they agree. Taking linear combinations of these Poisson brackets then results in Poisson brackets among possible expressions for the Hamiltonian constraint, as given in section \ref{Diracconst}.


\end{document}